\documentclass[a4paper]{article}

\usepackage{graphicx}

\usepackage{amssymb}

\newcommand{\be}{\begin{equation}}
\newcommand{\ee}{\end{equation}}
\def\e{{\rm e}}

\def\k{\kappa}

\begin{document}


\tolerance=10000

\def\pni{\par \noindent}
\def\vsh{\smallskip}
\def\s{\smallskip}
\def\vs{\medskip}
\def\vvs{\bigskip}
\def\vvvs{\bigskip\medskip} 
\def\vsp{\vsh \pni}
\def\vsn{\vsh\pni}
\def\cen{\centerline}
\def\ra{\item{a)\ }} \def\rb{\item{b)\ }}   \def\rc{\item{c)\ }}
\def\eg{{\it e.g.}\ } 
\hyphenpenalty=2000
\font\bfs=cmbx12 scaled\magstep1
\begin{center}
 FRACALMO PRE-PRINT  \  {\tt www.fracalmo.org}
 \\ {To appear in
Journal of Computational and Applied Mathematics (2008).} 
\vs

\hrule
\vvs

{\bfs{Evolution equations of the probabilistic}}
\vs

{\bfs {generalization of the Voigt profile function}}


\vvs

Gianni PAGNINI$^{(1)}$ and Francesco MAINARDI$^{(2)}$
\vs

$^{(1)}$
National Agency for  New Technologies,
 Energy and the Environment,\\ 
 ENEA, Centre  "E. Clementel", 
 \\ Via Martiri di Monte Sole 4,
 I-40129 Bologna, Italy\\
 {\tt gianni.pagnini@bologna.enea.it}
\vs

$^{(2)}$ Department of Physics, University of Bologna,
 and INFN, \\
   Via Irnerio 46, I-40126 Bologna, Italy\\
   {\tt francesco.mainardi@unibo.it}

\vs
{\small{Dedicated to Professor Jes\'us S. Dehesa on the occasion of his 60th
birthday}}
\\
 {\bf Revised Version, June 2008 }
\end{center}

\begin{abstract}
The spectrum profile that emerges in molecular spectroscopy and
atmospheric radiative transfer as the combined effect of Doppler and
pressure broadenings is known as the Voigt profile function.
Because of  its convolution integral representation,
the Voigt profile can be interpreted
as  the probability density function
of the sum of two independent random variables with
Gaussian density (due to the Doppler effect) and Lorentzian density
(due to the pressure effect).
Since these densities belong to the class of symmetric L\'evy
stable distributions,
 a probabilistic generalization is proposed as
the convolution of two arbitrary symmetric L\'evy densities.
We study the case when the widths of the considered distributions depend
on a scale-factor $\tau$ that is
representative of spatial inhomogeneity or temporal non-stationarity.
The evolution equations for
this  probabilistic generalization of the Voigt function are here introduced
and interpreted as generalized diffusion equations
containing two Riesz space-fractional derivatives,
thus classified as {\it space-fractional diffusion equations of double order}.
\end{abstract}

\vs


\vs


\section{Introduction}
In molecular spectroscopy and atmospheric radiative transfer,
the combined effects of Doppler and pressure broadenings
lead to the Voigt profile function, which 
turns out to be the convolution of the Gaussian (due to the Doppler broadening)
and the Lorentzian (due to the pressure broadening) distributions.
The study of the Voigt profile is an old issue in literature and 
many efforts have been directed to 
analyze its mathematical properties and relations with other special functions
and to obtain its numerical computation,
e.g. \cite{%
andersen-jqsrt-1978,armstrong-jqsrt-1967,%
dirocco_etal-as-2001,%
dulov_etal-jqsrt-2007,%
letchworth_etal-jqsrt-2007,%
mendenhall-jqsrt-2007,%
pagnini-mb-xxxx,%
sampoorna_etal-jqsrt-2007,%
yang-ijmest-1994,%
zaghloul-mnras-2007}.
\newpage
\vsp
In most papers appeared in the literature, the Voigt profile is defined in terms of a single weight-parameter,
that is the ratio of the Lorentzian to Gaussian width.
In physical applications, this parameter indicates which distribution is more important
than the other. 
Generally, the weight-parameter is considered with a constant value fixed by the
process.
Here we are interested to study the Voigt profile function when the
widths are not constant but depending on a scale-factor
with a power law. Physically, the one-dimensional variable of the Voigt function
is a wave-number and then this permits to take into account 
spatial inhomogeneity or
temporal non-stationarity when the scale-factor is the distance from an origin
or the elapsed time from an initial instant, respectively.
 \vsp
Since in probability theory the Gaussian and the Lorentzian distributions 
are known  to belong to the class of symmetric L\'evy stable distributions,
in this framework we propose   to generalize the Voigt function 
by adopting the convolution of two arbitrary symmetric L\'evy distributions.
Moreover, 
we provide the integro-differential equations
with respect to the scale-factor satisfied by the generalized
Voigt profiles. These evolution  equations can be interpreted as
space-fractional diffusion equations of double order.
\vsp
The  paper is organized as follows.
In section 2 the basic definitions for the Voigt profile are given.
In section 3 
the connection 
with the class of L\'evy stable distributions and 
the probabilistic generalization are introduced.
In section 4 we derive the integro-differential evolution equations of the 
generalized Voigt function with respect to the scale-factor. 
In section 5 the limits of low and high scale-factor values are investigated.
Finally, section 6 is devoted to the concluding remarks. 

\section{The Voigt profile function}
The Gaussian $G(x)$ and the Lorentzian $N(x)$ profiles 
are defined as
\be
G(x)=\frac{1}{\sqrt{\pi} \omega_G} \, \exp\left[-
\left(\frac{x}{\omega_G}\right)^2 \right] \,,
\quad 
N(x)=\frac{1}{\pi \omega_L} \, \frac{\omega_L^2}{x^2 + \omega_L^2} \,,
\label{definizioni}
\ee
where $\omega_G$ and $\omega_L$ are the corresponding widths.
From their convolution we have
the ordinary Voigt profile $V(x)$
\be
V(x)=\int_{-\infty}^{+\infty} N(x-\xi) G(\xi) \, d\xi = 
\frac{\omega_L/\omega_G}{\pi^{3/2}}
\int_{-\infty}^{+\infty}
\frac{\e^{-(\xi/\omega_G)^2}}
{(x-\xi)^2 + \omega_L^2} \, d\xi \,.
\label{voigt}
\ee
\begin{figure} 
\begin{center}
 \includegraphics[width=9cm,angle=-90]{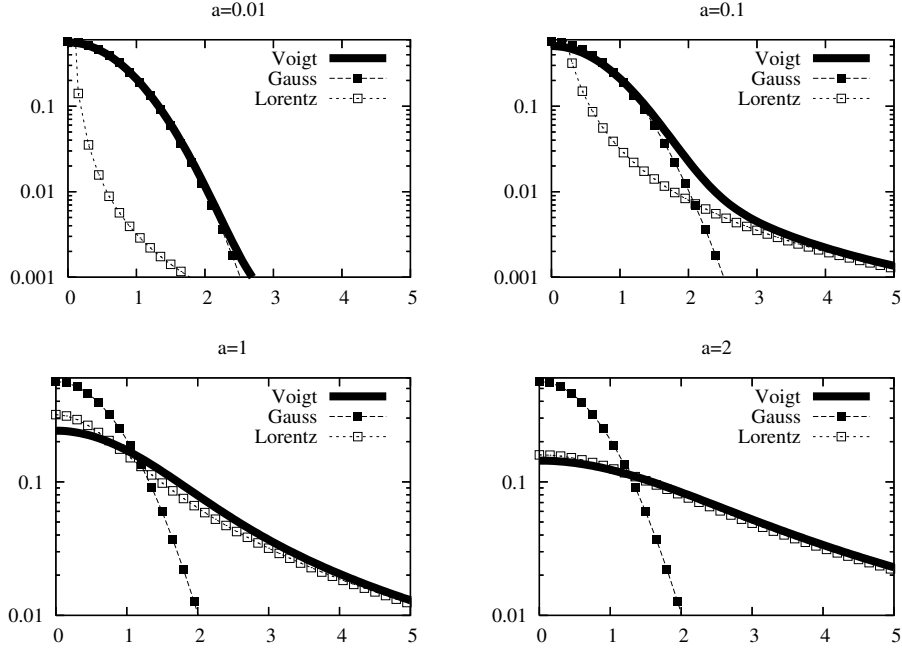}
\end{center}
\vskip-0.5truecm
\caption{Comparison between the Voigt, Gauss and Lorentz distributions with 
weight-parameter $a=\omega_L/\omega_G=0.01, 0.1, 1, 2$.}
\label{figura1}
\end{figure}
\vsp
The main parameter of the Voigt function is the weight-parameter $a$ defined as
$a=\omega_L/\omega_G$,
which is the ratio of the Lorentzian to Gaussian width and then
a measure of the relative importance between their influence on the
properties of the process. Generally, the $a<1$ case is important in astrophysics while
$a>1$ in spectroscopy of cold and dense plasma \cite{dirocco_etal-as-2001}.
In particular, two limits can be considered:
$i)$ $a \to 0$; $ii)$ $a \to \infty$.
In the first case, the
Lorentzian contribution is negligible in respect of the Gaussian one,
while the inverse occurs in the second case, see Fig. \ref{figura1}.
\vsp
Let $\widehat{f}(\k)$ be the characteristic function,
which is the Fourier transform of $f(x)$, so that
\be
\widehat{f}(\k)=\int_{-\infty}^{+\infty} \e^{+i \k x} f(x) \, d x \,,
\quad 
f(x)=\frac{1}{2\pi} \int_{-\infty}^{+\infty} \e^{-i \k x} \widehat{f}(\k) \, d\k \,,
\label{fourier}
\ee
then
$\widehat{V}(\k)=\widehat{G}(\k) \widehat{N}(\k)= 
\e^{-\omega_G^2 \k^2/4} \,
\e^{-\omega_L|\k|}
$.
These formulae imply the following integral representation
for the Voigt profile
\be
\! V(x) \!`= \! \frac{1}{2\pi} \int_{-\infty}^{+\infty}	
\!\! \e^{-i \k x} \e^{-\omega_G^2 \k^2 /4 - \omega_L |\k|} \, d \k 
= \frac{1}{\pi} \int_0^{\infty} 
\!\! \e^{-\omega_L \k - \omega_G^2 \k^2/4} \cos(\k x)\, d \k .
\label{cos-fourier-V}
\ee
\section{The probabilistic generalization of the Voigt profile function}
It is well known that if $X_1$ and $X_2$ are two independent random variables with
probability density function (PDF) $q_1$ and $q_2$, respectively, then the PDF
$p(w)$ of the random variable $W=X_1+X_2$ is given by the convolution integral
\be
p(w)=\int_{-\infty}^{+\infty} q_1(w-x_2) q_2(x_2) \, d x_2 \,.
\label{prob}
\ee
From (\ref{voigt}) and (\ref{prob}), 
the Voigt profile can be seen as the resulting PDF of the sum of
two independent random variables, one with Gaussian PDF and the other with Lorentzian PDF.
\vsp
The Voigt function has been generalized in literature in different ways, e.g. 
in \cite{sampoorna_etal-jqsrt-2007} the integrand function in formula (\ref{cos-fourier-V}) is
multiplied for a polynomial,  
in \cite{srivastava_etal-ass-1987,srivastava_etal-ass-1992,khan_etal-amc-2006} 
the cosine function in (\ref{cos-fourier-V}) 
is replaced with the Bessel function and
with the Wright function with one and more variables.
In this paper we propose a probabilistic generalization
in the framework of L\'evy distributions.
It is well known that the Gaussian and the Lorentzian distributions are 
two special cases of the class $\{L_\alpha(x)\}$ 
of the symmetric L\'evy stable distributions, 
where $\alpha$, $0 < \alpha \le 2$, is called characteristic exponent.
Writing 
the characteristic function (\ref{fourier}) of L\'evy distributions as
$\widehat{L}_\alpha(\k)= \e^{-|\k|^\alpha}$, 
the Gaussian and the Lorentzian profiles defined in (\ref{definizioni}) 
are recovered with $\alpha=2$ and $\omega_G=2$ and with $\alpha=1$
and $\omega_L=1$,
respectively,
\be
G(x)=L_2(x)=\frac{1}{2\sqrt{\pi}} \e^{-x^2/4} \,, \quad
N(x)=L_1(x)=\frac{1}{\pi}\frac{1}{x^2+1} \,.
\ee
The Voigt function can be straightforwardly generalized in probabilistic sense
by considering the sum of two independent random variables with
symmetric stable densities. Mathematically, this corresponds to
the convolution of two arbitrary symmetric L\'evy densities
of characteristic exponents 
$\alpha_1$ and $\alpha_2$. Denoting with $\mathcal{V}(x)$ the
generalized Voigt function, its integral representation 
and its characteristic function $\widehat{\mathcal{V}}(\k)$ are
\be
\mathcal{V}(x)=
\int_{-\infty}^{+\infty} L_{\alpha_1}(x-\xi) L_{\alpha_2}(\xi) \, d\xi \,,
\quad
\widehat{\mathcal{V}}(\k)=\e^{-|\k|^{\alpha_1} - |\k|^{\alpha_2}} \,.
\label{Vgen}
\ee

\section{The evolution equations with respect to the scale-factor}
\begin{figure}
\begin{center}
\includegraphics[width=9.0cm,angle=-90]{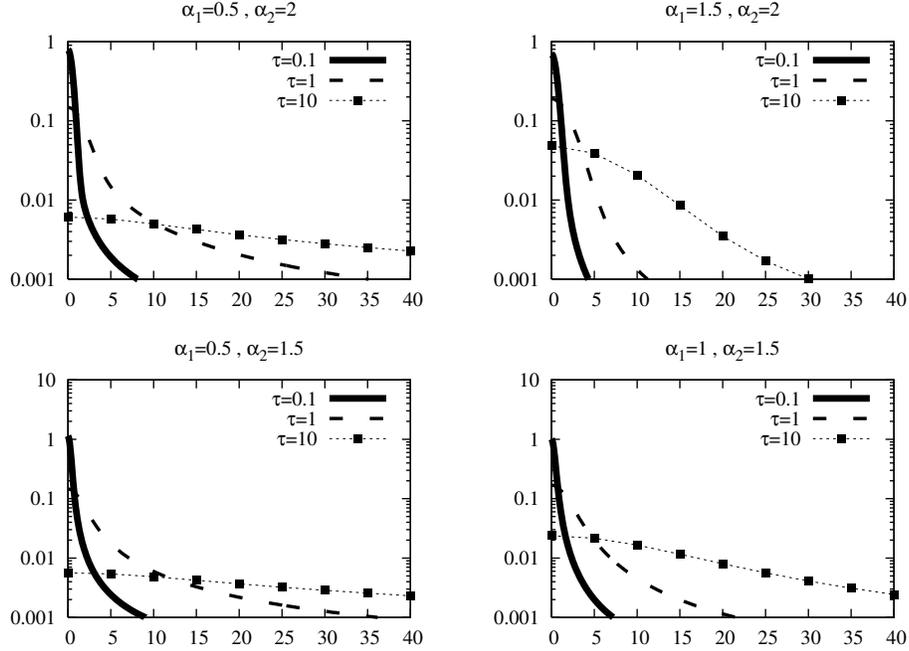}
\end{center}
\vskip -0.5 truecm
\caption{
The evolution of the generalized Voigt function $\mathcal{V}(x,\tau)$ for the pairs
$(\alpha_1,\alpha_2)=\{(0.5,2), (1.5,2), (0.5,1.5), (1,1.5)\}$
with $\tau=0.1,1,10$.
}
\label{figura2}
\end{figure}

In the previous sections the widths $\omega_G$ and $\omega_L$ are considered to be constants
and the weight-parameter $a$ fixed. However, differently from the previous papers on the topic,
we would like to know what happens when the widths $\omega_G$ and $\omega_L$ change
in space or time with a power law with respect to a scale-factor.
This analysis is relevant in inhomogeneous or not stationary cases,
for which the scale-factor
corresponds to the distance from an origin or the elapsed time, respectively.
Conversely, constant values of widths can be considered for homogeneous and
stationary cases.
In the present section we consider a scale-factor
$\tau$ for both spatial inhomogeneity and temporal non-stationarity.
\vsp
It is well known that
the L\'evy density functions $L_\alpha(x,\tau)$
are the fundamental solutions
of the space-fractional diffusion equation 
\cite{chechkin_etal-pre-2002,lumapa,sokolov_etal-appb-2004} 
\be
\frac{\partial L_\alpha(x,\tau)}{\partial \tau}={D_x^\alpha} L_\alpha(x,\tau) \,,
\quad L_\alpha(x,0)=\delta(x) \,, \quad 0 < \alpha \le 2 \,,
\label{sfde}
\ee
where ${D_x^\alpha}$ is a pseudo-differential operator known as
the Riesz space-fractional derivative of order $\alpha$.
Such pseudo-differential operator  
is defined in terms of its symbol $-|\k|^\alpha$,
i.e. the Fourier transform of $D_x^\alpha f(x)$ is $-|\k|^\alpha \widehat{f}(\k)$. 
We recall the explicit representations for $\alpha \ne 2$
\be
D_x^\alpha f(x)\!=\!\left\{
\begin{array}{lr}
\!{\displaystyle \frac{\Gamma(1+\alpha)}{\pi}\sin\left(\frac{\alpha \pi}{2}\right)
\! \int_0^{+\infty} \!\!\frac{f(x+\xi)-2f(x)+f(x-\xi)}{\xi^{1+\alpha}} \,d\xi \,,} & \alpha \ne 1 \\
\\
\!{\displaystyle -\frac{1}{\pi} \frac{d}{dx} \int_{-\infty}^{+\infty} \frac{f(\xi)}{x-\xi}\, d\xi} \,, & \alpha = 1
\end{array}
\right.
\ee
and the limit $D_x^{\alpha}f(x)=d^2 f/dx^2$ when $\alpha=2$.
The solutions of (\ref{sfde}) have the power law scaling property
\be
L_\alpha(x,\tau)=\tau^{-1/\alpha} L_\alpha \left(\frac{x}{\tau^{1/\alpha}}\right) \,.
\label{scaling}
\ee
For analytical and graphical representations of stable densities we
refer the reader to \cite{lumapa}.
We observe that the distributions defined in (\ref{scaling})
are self-similar and they obey to the same power law scaling for all values
of the scale-factor $\tau$. In particular, for $\alpha=2$ and $\alpha=1$
the Gaussian and the Lorentzian densities are recovered, respectively, 
and from (\ref{scaling}) we have 
$\omega_G \propto \tau^{1/2}$ and 
$\omega_L \propto \tau$.
\vsp
Following scaling (\ref{scaling}),
our generalized Voigt function (\ref{Vgen}) becomes
\be
\mathcal{V}(x,\tau)=\tau^{-1/\alpha_1-1/\alpha_2} \int_{-\infty}^{+\infty}
L_{\alpha_1}\left(\frac{x-\xi}{\tau^{1/{\alpha_1}}}\right)
L_{\alpha_2}\left(\frac{\xi}{\tau^{1/{\alpha_2}}}\right) d\xi \,,
\label{voigtgen}
\ee
and its characteristic function is
\be
\widehat{\mathcal{V}}(\k,\tau)=\e^{-|\k|^{\alpha_1}\tau-|\k|^{\alpha_2}\tau} \,.
\label{fourier2}
\ee
In this case, it is possible to show that formula (\ref{voigtgen}) is the solution of
the following integro-differential equation
\be
\frac{\partial \mathcal{V}}{\partial \tau}=
{D_x^{\alpha_1}} \mathcal{V}(x,\tau) + {D_x^{\alpha_2}} \mathcal{V}(x,\tau) \,, \quad
\mathcal{V}(x,0)=\delta(x) \,.
\label{eq2}
\ee
In fact, after the Fourier transformation 
Eq. (\ref{eq2}) becomes
\be
\frac{\partial \widehat{\mathcal{V}}}{\partial \tau}=
- |\k|^{\alpha_1} \widehat{\mathcal{V}}(\k,\tau) 
- |\k|^{\alpha_2} \widehat{\mathcal{V}}(\k,\tau) \,,
\quad
\widehat{\mathcal{V}}(\k,0)=1 \,,
\ee
which is solved by (\ref{fourier2}).
When $\alpha_1=1$ and $\alpha_2=2$ 
the integro-differential equation (\ref{eq2}) is the evolution equation of 
the ordinary Voigt function.
Equation (\ref{eq2}) can be classified as  
{\it space-fractional diffusion
equation of double order}. 
The evolution of $\mathcal{V}(x,\tau)$ 
for different pairs of $(\alpha_1,\alpha_2)$
with $\tau=0.1,1,10$ is shown in
Fig. \ref{figura2}.


\section{The asymptotic scaling laws for low and high scale-factor}
The Voigt (\ref{voigt}) and the generalized Voigt (\ref{Vgen}) profiles are 
derived from the convolutions of two
self-similar processes with different scaling laws and,
as a consequence, the similarity is lost.
However, we ask which are the scaling laws of the Voigt functions
in the limits of low and high values of the scale-factor $\tau$.
 \vsp
Since for L\'evy stable densities with $\alpha \neq 2$ the mean square displacement diverges,
the same occurs
for the ordinary and the generalized Voigt functions. Then, to analyze the scaling laws
when $\tau \to 0$ and $\tau \to \infty$ the variance $\langle x^2 \rangle$
cannot be used. 
However, 
the  scaling law of a L\'evy stable process with characteristic exponent $\alpha$
can be studied by mean the quantity
$\langle |x|^q \rangle^{1/q}$, with $0< q < \alpha$. In this respect, we have
\be
\langle |x|^q \rangle= 2 \, \int_0^{+\infty} x^q \mathcal{V}(x,\tau) \, dx=
2 \, \mathcal{V}^*(q+1,\tau) \,,
\quad 0 < q < \min\{\alpha_1,\alpha_2\} \,,
\ee
where $\mathcal{V}^*(s)$ is the Mellin transform of $\mathcal{V}(x)$, $x > 0$,
defined as \cite{paris-kaminski}
\be
\mathcal{V}^*(s)=\int_0^{+\infty} \mathcal{V}(x) x^{s-1} \, dx \,, \quad \quad
\mathcal{V}(x)=\frac{1}{2 \pi i} \int_{c-\infty i}^{c+\infty i} \mathcal{V}^*(s) x^{-s} \, ds 
\,.
\ee
Without lost in generality, let us state $\alpha_1 < \alpha_2$.
Starting from (\ref{Vgen}) and following \cite{pagnini-mb-xxxx},
the Mellin-Barnes integral representation of  
the generalized Voigt function $\mathcal{V}(x)$ is
\begin{eqnarray}
\mathcal{V}(x,\tau) &=&
\frac{\tau^{-1/\alpha_2}}{\alpha_2 \pi}
\frac{1}{2 \pi i} \int_{\mathcal{L}_0} \,
\frac{1}{2 \pi i} \int_{\mathcal{L}_1}  
\Gamma(s_0) \Gamma(s_1)
\Gamma\left(\frac{1-s_0-\alpha_1 s_1}{\alpha_2}\right) \nonumber \\
& & \qquad 
\times \quad \tau^{s_0/\alpha_2 + (\alpha_1/\alpha_2-1)s_1}\cos(s_0 \pi/2)
x^{-s_0} \, ds_0 ds_1
\,.
\end{eqnarray}
\newpage
\noindent
Hence its Mellin transform is
\begin{eqnarray}
\mathcal{V}^*(s_1,\tau) &=&
\frac{\tau^{(s_0-1)/\alpha_2}}{\alpha_2 \pi}
\Gamma(s_0) \cos(s_0 \pi/2) \nonumber \\ 
& & \quad 
\times \; \frac{1}{2 \pi i} \int_{\mathcal{L}_1}  
\Gamma(s_1)
\Gamma\left(\frac{1-s_0-\alpha_1 s_1}{\alpha_2}\right)
\tau^{(\alpha_1/\alpha_2-1)s_1} ds_1
\,,
\end{eqnarray}
and finally,
\begin{eqnarray}
\langle |x|^q \rangle &=& 2 \mathcal{V}^*(q+1,\tau) \nonumber \\		
&=& - \frac{2 \, \tau^{q/\alpha_2}}{\alpha_2 \pi} \Gamma(q+1) \sin(q \pi/2) \nonumber \\
& &  \qquad  
\times \; \frac{1}{2 \pi i} \int_{\mathcal{L}_1} 
\Gamma(s_1)\Gamma\left(\frac{-q-\alpha_1 s_1}{\alpha_2}\right)
\tau^{(\alpha_1/\alpha_2-1)s_1} ds_1 \,.
\end{eqnarray}
Applying the residue theorem to
$\Gamma\left(\frac{-q-\alpha_1 s_1}{\alpha_2}\right)$, we obtain
the convergent series for $\tau \to \infty$,
\be
\langle |x|^q \rangle =
- \frac{2  \tau^{q/\alpha_1}}{\alpha_1 \pi} \Gamma(q+1) \sin(q \pi/2)
\sum_{n=0}^{\infty} \frac{(-1)^n}{n!}
\Gamma\left(\frac{\alpha_2 n -q}{\alpha_1}\right)
\tau^{-n(\alpha_2/\alpha_1-1)} \,.
\ee
Applying the residue theorem to $\Gamma(s_1)$, we obtain the convergent series for
$\tau \to~0$,
\be
\langle |x|^q \rangle =
- \frac{2  \tau^{q/\alpha_2}}{\alpha_2 \pi} \Gamma(q+1) \sin(q \pi/2)
\sum_{n=0}^{\infty} \frac{(-1)^n}{n!}
\Gamma\left(\frac{\alpha_1 n -q}{\alpha_2}\right)
\tau^{n(1-\alpha_1/\alpha_2)} \,.
\ee
In papers \cite{chechkin_etal-pre-2002,sokolov_etal-appb-2004} the 
limits $(\tau \to 0 \,, \tau \to \infty)$ are computed using a different method 
but with the same results.
Then the two limits under consideration give 
\be
\left\{
\begin{array}{lr}
\langle x^q \rangle^{1/q} \propto \tau^{1/\alpha_2} \,, & \tau \to 0\,; \\
\langle x^q \rangle^{1/q} \propto \tau^{1/\alpha_1} \,, & \tau \to \infty\,. \\
\end{array}
\right.
\ee
In the limit $\tau \to 0$, the corresponding scaling law of the generalized Voigt profile 
is governed by the
L\'evy density with the higher value of the characteristic exponent,
while in the limit $\tau \to \infty$ by that with the lower value.
In particular, for the ordinary Voigt profile
$(\alpha_1=1 \,, \alpha_2=2)$ the process scales as $\tau^{1/2}$ and $\tau$ for
law and high values of the scale-factor, respectively.
This means that if the power law represents 
inhomogeneity, or non-stationarity, the resulting profile is approximated by
a Gaussian for small distances from an origin, or small
elapsed times, and it is approximated by a Lorentzian for large distances, or
large elapsed times.
This result is consistent with the usual limits $a \to 0$ and
$a \to \infty$, see Fig. \ref{figura1}.
\newpage
\section{Conclusions}
In the present paper we have considered the Voigt profile function
and proposed a probabilistic generalization
as the convolution of two arbitrary symmetric L\'evy densities.
Generally, the Voigt profile characteristics are studied with respect to a weight-parameter $a$
that is the ratio of Lorentzian to Gaussian widths, $a=\omega_L/\omega_G$,
and it is assumed to be a constant property of the process.
Conversely, here we have considered both widths depending
on a scale-factor $\tau$ that is representative of inhomogeneity or 
non-stationarity.
We have introduced parametric integro-differential equations for the ordinary and
the generalized Voigt functions. These integro-differential equations can be classified
as {\it space-fractional diffusion equations of double order} because they include
two Riesz space-fractional derivative of different orders. 
In this respect, the present paper shows an application in physics of the 
distributed fractional derivatives formalism.
\vsp
Finally, the limits of the Voigt function for low and high values of the scale-factor 
are considered. 
The Voigt function turns out to be not self-similar, 
even if it is expressed as the convolution of two self-similar L\'evy processes.
Its scaling law
is dominated by the L\'evy density with the higher value of the characteristic 
exponent when $\tau \to 0$ and by that with the lower value 
when $\tau \to \infty$. 

\section*{Acknowledgment}
GP would like to thank GNFM-INdAM (Progetto Giovani 2007) for 
partial financial support to attend
the interdisciplinary conference in honor of 
Prof. Jes\'us S. Dehesa's 60th birthday 
``Special Functions, Information Theory and Mathematical Physics'',
held in Granada, Spain, September 17-19 2007.



\end{document}